\documentclass[twocolumn,aps]{revtex4}

\usepackage{amsfonts}
\usepackage{amsmath}
\usepackage{bm}
\usepackage{graphicx}
\usepackage{xcolor}
\usepackage{color}

\def \beq {\begin{equation}}
\def \eeq {\end{equation}}
\def \bea {\begin{eqnarray}}
\def \eea {\end{eqnarray}}
\def \bfig {\begin{figure}}
\def \efig {\end{figure}}

\def \lab {\label}
\def \bx {\bm{x}}

\def \bB {\bm{B}}

\def \bE {\bm{E}}
\def \bv {\bm{v}}
\def \bb {\bm{b}}
\def \bu {\bm{u}}

\def \bz {\bm{z}}
\def \bna {\bm{\nabla}}
\def \bE {\bm{E}}

\def \bj {\bm{j}}
\def \nn {\nonumber}
\def \de {\partial}
\def \le {\left}
\def \ri {\right}

\def \bx {\bm{x}}

\def \fr {\frac}

\begin{document}

\title{ Explosive Particle Dispersion in Plasma Turbulence }
\author{ S. Servidio$^{1}$,  C. T. Haynes$^2$, W. H. Matthaeus$^{3}$, D. Burgess$^2$, V. Carbone$^{1}$, and P. Veltri$^{1}$ }

\affiliation{
$^1$Dipartimento di Fisica, Universit\`a della Calabria, I-87036 Cosenza, Italy\\
$^2$School of Physics and Astronomy, Queen Mary University of London, Mile End Road, London E1 4NS, United Kingdom\\ 
$^3$Bartol Research Institute and Department of Physics and Astronomy, University of Delaware, Newark, DE 19716, USA }

\date{\today} 

\input epsf

\begin{abstract} 
Particle dynamics are investigated in plasma turbulence, 
using self-consistent kinetic simulations, in two dimensions. In steady state, the trajectories of 
single protons and proton-pairs are studied, at different values of plasma $\beta$ (ratio between kinetic 
and magnetic pressure). For single-particle displacements, results are consistent with fluids and magnetic 
field line dynamics, where particles undergo normal diffusion for very long times, with higher $\beta$'s being more diffusive.
In an intermediate time range, with separations lying in the inertial range, particles experience an explosive 
dispersion in time, consistent with the Richardson prediction.
   These results, obtained for the first time with a self-consistent 
   kinetic model, are relevant for
astrophysical and laboratory plasmas, where turbulence is crucial for heating, 
mixing and acceleration processes.
\end{abstract}

\maketitle 

  The motion of particles in complex fields has been one of the most 
  fascinating problems in physics, with interdisciplinary applications that span from 
  hydrodynamics to astrophysical plasmas. The study of Lagrangian tracers 
  is complementary to the theory of turbulence \cite{Kolmogorov41a} wherein
  individual tracers undergo a random motion, asymptotically approaching the 
  diffusive Brownian behavior \cite{Langevin08}. The relative motion of a pair of 
  tracers is a different and more subtle problem, as the growth of separation may 
  reflect turbulent correlations \cite{Richardson26}. Both individual and pair 
  particle transport are of great importance in applications ranging from 
  laboratory plasmas \cite{TaylorMcNamara71,HauffEA09} to  magnetic field wandering 
  and tangling in the galaxy \cite{JokipiiParker69,EyinkEA13,LazarianEA15}, 
  corona \cite{LepretiEA12} and interplanetary medium \cite{RuffoloEA03,RuffoloEA04}.
  Often discussed in the purely diffusive limit, these varieties of transport may 
  also frequently display nondiffusive (superdiffusive or subdiffusive) behavior 
  (e.g., \cite{PerroneEA13}). These subjects have been studied mainly in the 
  test-particle approximation, appropriate, for example, in describing high energy 
  cosmic  rays \cite{Jokipii66}.

  When the transported particles are elements of the thermal 
  plasma \cite{TaylorMcNamara71,Balescu94}, the distribution is often taken 
  as an equilibrium Maxwellian. In this context, test-particles and passive tracers 
  in Magnetohydrodynamics (MHD) have been of 
  interest \cite{DmitrukEA04,RuffoloEA04,ZimbardoEA06,BusseEA07}. 
  However, for low collisonality plasmas where kinetic effects typically generate strong 
  departures from thermal Maxwellian equilibria \cite{Marsch06}, one should treat 
  the transport problem self-consistently. We present first results on this 
  fundamental topic in the present Letter.

In the case of stationary random motion, 
a single fluid element at position ${\bm x}(t)$ and velocity ${\bm v(t)}$ has a finite 
auto-correlation time (or Lagrangian integral time) 
\beq
\tau_\ell = \fr{1}{\langle {v(t_0)}^2\rangle} \int_0^{\infty} \langle {\bm v}(t_0)\cdot{\bm v}(t_0+\tau)\rangle d \tau =  \fr{D_s}{\langle v^2\rangle}, 
\lab{eq:tc}
\eeq
where the ensemble $\langle \bullet \rangle$ has been computed over a large number of realizations, 
positions and times, and $D_s$ is the diffusion coefficient. 
The mean square displacement of $\Delta{\bm s}={\bm x}(t_0+\tau)-{\bm x}(t_0)$, 
in the limit of $\tau\gg \tau_\ell$, obeys 
\beq
\langle \Delta {s}^2 \rangle = 2 D_s \tau. 
\lab{eq:ds}
\eeq
The above represents the long-time limit diffusive behavior, typical of Brownian motion. 
In the opposite limit, $\tau\rightarrow 0$, in the so-called dissipative range, 
particles conform to ballistic transport, governed by $\langle \Delta {s}^2 \rangle\sim \tau^2$ \citep{Batchelor50,FalkovichEA01}.

Together with the asymptotic behavior of single particle motion, it is interesting to consider the
motion of two particles, as done by \citet{Richardson26}. In this pioneering work
it was predicted that, at intermediate separations,
the inner-particle distance $r^2\equiv|{\bm r}_{1,2}|^2=|{\bm x}_1(\tau)-{\bm x}_2(\tau)|^2$
is super-diffusive in time. Averaging over time and volume, it is observed that 
\beq
\langle r^2 \rangle \sim \tau^3.
\lab{eq:r2}
\eeq
This motion is very rapid, explosive in time, and is related to the mixing properties of a
turbulent field. Richardson obtained this law from basic principles, computing solutions 
to the particle-pairs probability distribution, and using hints from observations. 
Note that this work has been a precursor of Kolmogorov theory of turbulence, 
and here will be applied to  kinetic self-consistent models of plasmas.

Single particle displacement and pair dispersion are here investigated in plasmas, 
using self-consistent kinetic models of turbulence \cite{ServidioEA12,HaynesEA14,FranciEA15}. 
We study the motion of the plasma particles themselves, represented by 
elements of the proton distribution function, in the phase-space given by position and velocity.
We will emphasize a novel study of the particle statistics in a collisionless plasma, 
in a driven turbulent state, for different plasma parameters.


Driven simulations of the hybrid-PIC model (kinetic ions and fluid electrons) have been performed 
 (ions hereafter are intended to be protons), 
in a 2.5D geometry, solving \citep{Winske85,Matthews94}
\bea
\!\!\!\!&&\fr{\de \bx}{\de t} = \bv,~~\fr{\de \bv}{\de t} = \bE + \bv\times\bB,\nn \\
\!\!\!\!&&\fr{\de \bB}{\de t}\!=\!-\bna\!\times\!\bE\!=\!\!\bna\!\times\!\!\!\le[
\bu\!\times\!\bB\!-\!\fr{1}{\rho}\bj\!\times\!\bB\!+\!\!\fr{1}{\rho}\bna P_e\!\!-\!\!\eta\bj
\!\ri]\!,
\lab{eq:pic}
\eea
where $\bx$ are the proton positions and $\bv$ their velocities, $\bB=\bb+B_{0}{\hat \bz}$ 
is the total (solenoidal) magnetic field, $\bj=\bna\times\bB$ is the current density, 
$\rho$ and $\bu$ represent the proton (electron) density and the proton bulk velocity, respectively. 
 Electron pressure $P_e$ is adiabatic, with $\beta_e=\beta_p=\beta$, 
and a small resistivity $\eta$ suppresses small grid-scale activity. 
Space is normalized to the proton skin depth $d_p$, time with the 
proton cyclotron frequency $\Omega_{cp}$, velocities to the thermal speed $v_{th}$, 
and magnetic field with Alfv\`en speed of the mean magnetic field $B_0$.
A spatial grid of $N_x\times N_y=512^2$ mesh points is defined in a periodic 
box of side $L_0=128d_p$. A large number (1500) of particles-per-cell (ppc) has been chosen 
to suppress the statistical noise.
Three values of plasma $\beta$ (thermal/magnetic pressure)
are chosen, as reported in Table \ref{tabr}.
The initial fluctuations are chosen with random 
phases, and with the Fourier modes satisfying $3\leq m \leq 7$, 
where the $k$-vector is defined as $k=\fr{2\pi}{128 d_p}m$. 
Fluctuations have $b_{rms}=v_{rms}=0.5$, with $B_{0}=1$. 
Proton heating in low-noise simulations is moderate \cite{FranciEA15}, and the value of the effective  $\beta$ at the end of each simulation is increased by $\sim12\%$.

\begin{table}
  \begin{center}
\def~{\hphantom{0}}
  \begin{tabular}{ l | l | l | l | l | l | l | l | l }
    \hline
    ~   &     $\beta$  &   $\gamma$   &  $\tau_{\ell}$ &  $\tau_{\ell g}$ & $D_s$  & $D_s^{(a)}$ & $\mu$ & $\chi_0$ \\
    \hline 
    \hline
    Run I     & $0.1$  &   $1.07$     &   $11$  &   $22$      & $2.66$  & $2.56$  & $1.97$ & $0.11$  \\
    Run II    & $0.5$  &   $1.07$     &   $5$   &   $17$      & $2.77$   & $2.65$  & $1.99$ & $0.15$  \\
    Run III   & $5.0$  &   $1.21$     &   $1$   &   $7$       & $3.64$  & $3.64$  & $1.80$ &  $0.47$  \\
    \hline
  \end{tabular}
  \caption{
Plasma $\beta$; structure function exponent $\gamma$; 
Lagrangian integral times ($\tau_\ell$ and $\tau_{\ell g}$, in units of $\Omega_{cp}^{-1}$); diffusion coefficient $D_s$
and its expectation $D_s^{(a)}$;  pair-dispersion exponent $\mu$ and pair-diffusion coefficient $\chi_0$.
}
  \label{tabr}
  \end{center}
\end{table}

To achieve steady state turbulence in a plasma, we borrow ideas from 
hydrodynamics \cite{ChenEA93,BecEA10,ThalabardEA14}. 
We initially let the system decay freely, and then we introduce
 a forcing at the peak of nonlinearity $t_\star$
(roughly the peak of $\langle j_z^2 \rangle$ \cite{Mininni09}), 
with $t_\star\sim 25\Omega_{cp}^{-1}$. The 
forcing consists of ``freezing'' the amplitude of the large scale 
modes of the inplane magnetic field, with $1\leq m \leq 4$, leaving the 
phases unchanged. This corresponds to a 
large scale input of energy. We perform the analysis 
described below when a steady state has been achieved,
for $50<t\Omega_{cp}<250$.

To characterize turbulence, we computed the second order structure function of the magnetic field 
$S_b({\bm\delta})=\langle\left[{\bm b}\left({\bm x}+{\bm \delta}, t\right)-{\bm b}\left({\bm x}, t\right)\right]^2\rangle_{V, T}$, 
where $\langle \bullet \rangle_{V, T}$ represents a double average over volume and time.
Positions $\bm x$ and increments $\bm \delta$ are in the $(x,y)$ plane.
For an isotropic inertial range of turbulence,
\beq
S_b(\delta) \sim \delta^{\gamma}. 
\lab{eq:s2gamma}
\eeq
As reported in Fig.  \ref{fig:ebsb}, the structure function manifests a clear self-similar range.
Fitting with Eq.~(\ref{eq:s2gamma}), we find
that $\gamma$ is quite close to unity, as reported in the Table \ref{tabr}.
Note that in classical 3D hydrodynamic turbulence at large Reynolds number, 
$\gamma=2/3$, corresponding 
to the celebrated Kolmogorov law \cite{Kolmogorov41a}. 
In plasmas the case is more complex, 
and it can depend on other factors, 
such as compressibility, dimensionality and anisotropy, 
as well as the effective Reynolds numbers.
Note, however, that observations and simulations suggest non-universality 
of plasma turbulence \cite{LeeEA10,TesseinEA09}.

\bfig
\epsfxsize=8cm
\centerline{\epsffile{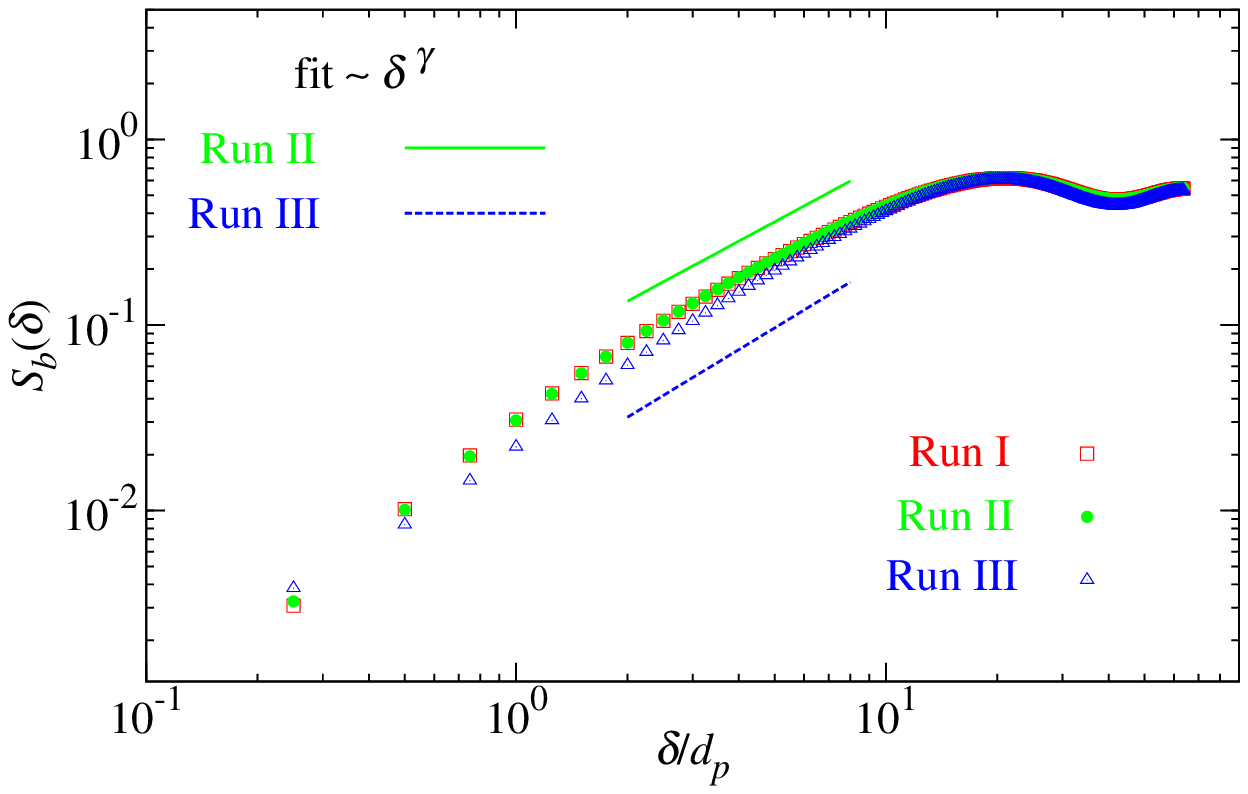}} 
\caption{Structure function of the magnetic field as a 
function of the spatial increment $\delta$, for all the Runs. 
Full (green) and dashed (blue) lines represent the fit with Eq.~(\ref{eq:s2gamma}), 
for Run II and III, respectively. Exponents $\gamma$ are given in Table \ref{tabr}.}
\lab{fig:ebsb}
\efig

We computed the auto-correlation function $C_b({\bm \delta})=\langle{\bm b}\left({\bm x}+{\bm \delta}, t\right)\cdot{\bm b}\left({\bm x}, t\right)\rangle_{V, T}$, 
and the auto-correlation length as $\lambda_C =\int_0^{L_0/2} C_b(q) dq$. 
For these simulations $\lambda_C\sim 9d_p$, 
which provides a large scale bound to Eq.~(\ref{eq:s2gamma}). 
Analogously, one might 
identify the small scale termination of the inertial range 
approximately as the Taylor microscale, which in our case is
$\lambda_T=\sqrt{\langle b^2\rangle/\langle j^2\rangle}\sim1.5 d_p$. 
>From Fig.  \ref{fig:ebsb}, Eq.~(\ref{eq:s2gamma}) holds for $\lambda_T<\delta<\lambda_C$.
It is interesting to note that, at the highest $\beta$ (Run III), a slightly shorter 
inertial range is observed, with an higher value of $\gamma$. This is possibly 
due to an higher damping of the Alfv\'enic and magnetosonic activity.

We analyzed a subset of $N_p$ randomly selected particles, 
represented by particle-in-cell pseudo-particles, with $N_p=10^5$. Convergence tests
have been performed varying $N_p$ from $5\times10^4$ to $N_p=2\times10^5$ showing no significant difference.
The space-time trajectories of some ``puffs'' of particles, located at
different (randomly selected) regions, are reported 
in Fig.  \ref{fig:puffs}. In the same plot, shaded contours reports 
$j_z$ at $t\Omega_{cp}\sim 50$ and $250$.
Particles bunches spread explosively in time, with a very fast departure in the 
first 10-30 cyclotron times. The inset shows the initial spreading of the 
central puff, together with some the trajectories of the associated gyro-centers. 
Gyro-center positions have been computed as ${\bm x}_g(t)=(1/T)\int_{t-T/2}^{t+T/2} {\bm x}(t')dt'$, 
using the gyroperiod $T=2\pi{\Omega_{cp}^{-1}}$.
The initial separation suggests a superdiffusive behavior, while at very long times the motion seems to be uncorrelated and erratic.
Trajectories vary greatly: some remain near the origin; others experience long flights; some rapidly change direction.  
These differences may reveal interesting correlations between particles and local structures \cite{RuffoloEA03,TooprakaiEA07,DrakeEA10,ZankEA15}, 
and will be matter of future investigations. 
As mentioned above, some trajectories are similar to test particles in MHD or model fields \cite{SeripienlertEA10}.

\bfig
\epsfxsize=8.3cm
\centerline{\epsffile{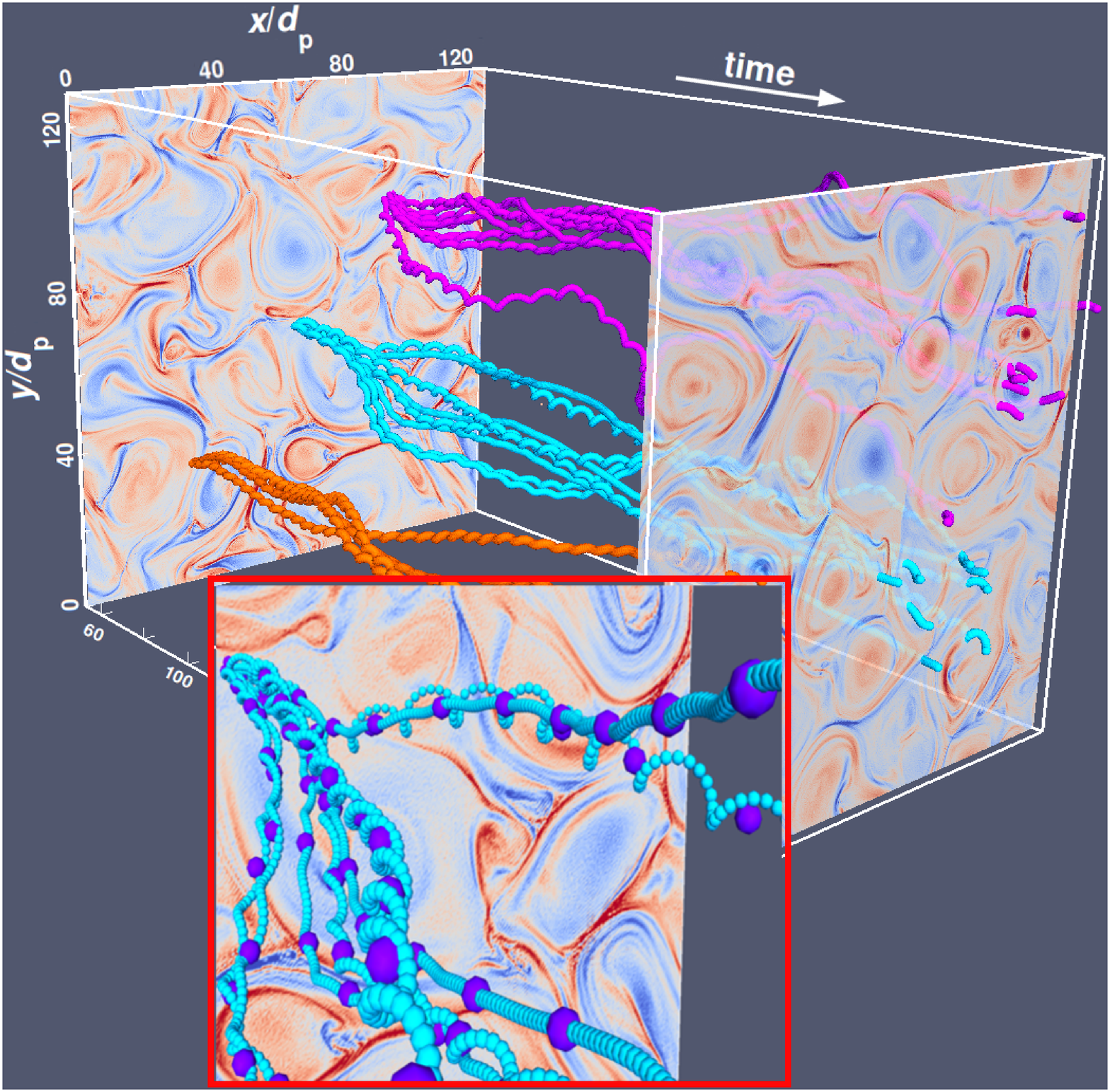}} 
\caption{ 
``Puffs'' of particles as a function of time, starting from the steady state, located at different regions of 2D  plasma turbulence. $j_z$ is shown at two times, $t\Omega_{cp}=50, 250$ (shaded surfaces). The inset shows the initial spreading of a population (small subset), up to $t\Omega_{cp}\sim75$ (small spheres), together with the position of some gyrocenters (big spheres). Explosive dispersion is observed.
}
\lab{fig:puffs}
\efig

To understand the ergodic motion in Fig. \ref{fig:puffs}, we 
analyzed single-particle statistics. We computed the Lagrangian correlation times defined by 
Eq.~(\ref{eq:tc}), for both particles ${\bm x}(t)$ and gyro-centers ${\bm x}_g(t)$.
These correlation times $\tau_\ell$ and $\tau_{\ell g}$, reported in the Table for all runs, 
are larger than the cyclotron time and depend on the value of $\beta$, being much smaller for 
higher $\beta$'s. This faster decorrelation is evidently due to the higher plasma thermal noise
which decorrelates the motion earlier. The gyrocenters have longer decorrelation times.

To establish the link between plasma particles and fluid tracers, we analyzed the 
single-particle displacement $\langle {\Delta s}^2 \rangle$.  Here brackets indicate again an average 
over particles and times. Following Eq.~(\ref{eq:ds}), one can compute the 
running diffusion coefficient as 
$D_s = \fr{1}{2} \fr{\de \langle{{\Delta{s}^2}}\rangle}{\de \tau}$.
If the displacement is stochastic, for times $\tau\gg \tau_\ell$, $D_s \rightarrow const.$
On the contrary, for $\tau \rightarrow 0$,  $\langle {\Delta s}^2 \rangle\sim \tau^2$, typical of ballistic transport. 
As reported in Fig.~\ref{fig:diff}, 
$\langle\Delta s^2\rangle$ behaves asymptotically as $\sim \tau$.
The horizontal lines indicating $D_s$ computed as a fit for very large times, namely $\tau>90\Omega_{cp}^{-1}\gg\tau_\ell$.
This value can be compared with the asymptotic coefficient, computed from Eq.~(\ref{eq:tc}) 
as $D_s^{(a)}=\int_0^\infty \langle {\bm v}(t_0)\cdot{\bm v}(t_0+\tau)\rangle d\tau$.
As it can be seen, from the figure and the Table, the long-time diffusive limit is evident \citep{TaylorMcNamara71,OkudaDawson73}. 
As expected from the Lagrangian correlation times estimation, 
plasmas with higher $\beta$ (Run III) are more diffusive, 
with the decorrelation being faster, 
due to the enhanced importance of fast microscopic particle speeds.  
(Note that the typical oscillation of running diffusion coefficients, 
commonly observed in test-particle studies,  
have a period on the order of the cyclotron time.)
In the inset of Fig.~\ref{fig:diff}, 
the mean square displacement is shown at earlier times, 
for Run II (all the runs have similar behavior, not shown here). It is evident that the Batchelor regime, 
where $\sim\tau^2$ \citep{FalkovichEA01}, is observed for $\tau<1.7\Omega_{cp}^{-1}\ll\tau_\ell$.

\bfig
\epsfxsize=8.6cm
\centerline{\epsffile{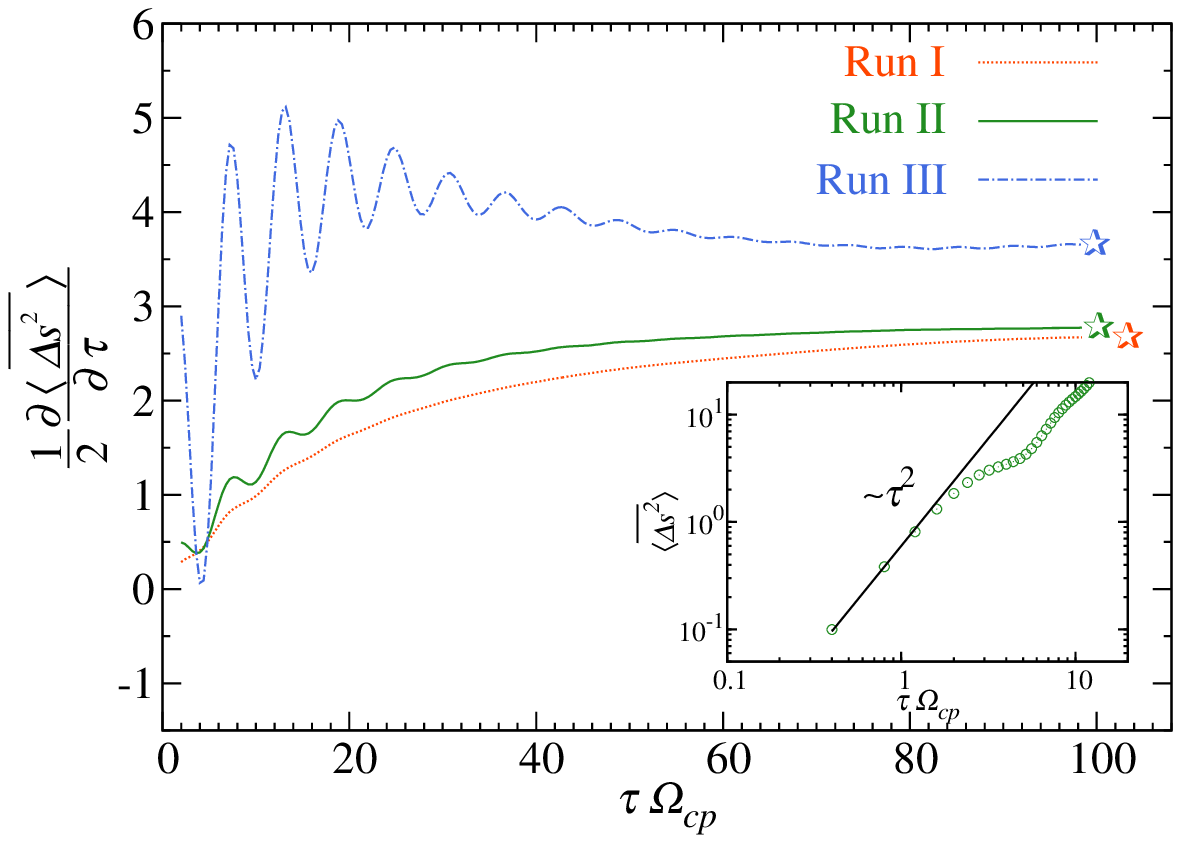}} 
\caption{Running diffusion coefficients $\frac{1}{2}\frac{\partial \langle {\Delta s^2} \rangle}{\partial \tau}$
for all the runs. Stars indicate $D_s$, 
computed as a fit for very large times $\tau\Omega_{cp}>90$.
The inset shows the mean square displacement $\langle\Delta s^2\rangle$ in the very initial stage (Run II).}
\lab{fig:diff}
\efig

\bfig
\epsfxsize=8.8cm
\centerline{\epsffile{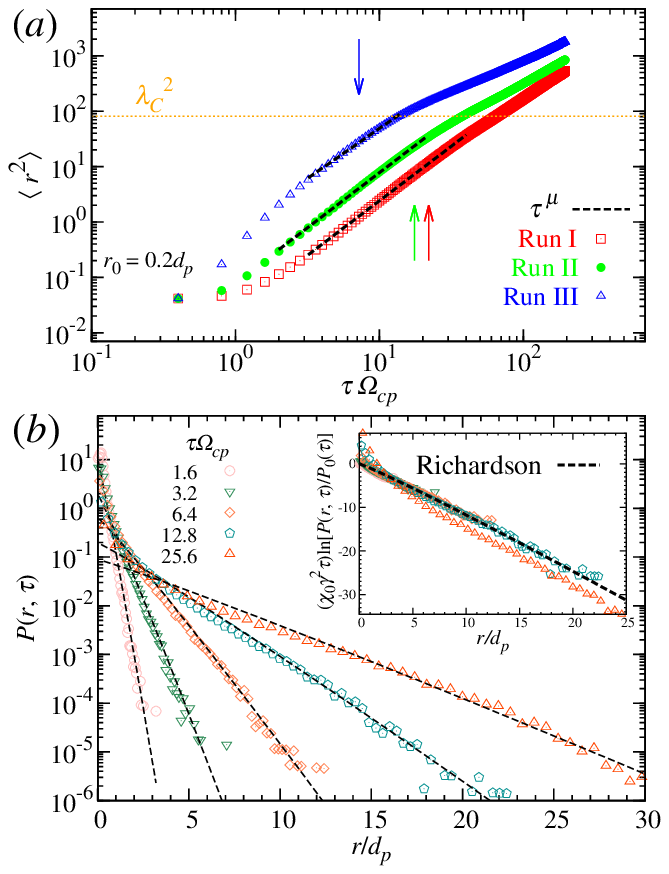}} 
\caption{
(a) Mean squared gyro-centers separation as a function of time. 
Inertial range fits $\sim \tau^\mu$ are reported (black dashed lines) (see Table I for $\mu$.) 
The horizontal (orange) dotted line represents $\lambda_C$, while arrows indicate
$\tau_{\ell g}$. 
(b) Particle separation probability $P(r, \tau)$ as a function of $r$, at different $\tau$. 
Results are shown for the intermediate $\beta$ (Run II) but are similar for all the runs. 
The Richardson fit is reported with dashed (black) lines. 
Inset of (b): rescaled $P(r, \tau)$ for the same times (symbols) and Richardson expectation ($-r^\gamma$) (dashed line). 
The generalized law is clearly observed for $t\sim\tau_{\ell g}$ , 
being lost when $t\gg\tau_{\ell g}$ (red triangles).
}
\lab{fig:r2}
\efig

For times on the order of $\tau_\ell$ and $\tau_{\ell g}$, an 
interesting transient is observed, resembling the super-diffusive behavior typical of fluids.
These time ranges correspond to the fast dispersive motion observed in Fig.~\ref{fig:puffs}.
We study the temporal behavior of gyrocenters distances $r(\tau)$ (in order to avoid 
the trivial particle gyroperiod), randomly selected in our system, 
where the initial separation $r_0$ has been chosen to be sufficiently small.
In ordinary fluids, 
in order to capture inertial range super-diffusion, 
this separation 
must fall in the dissipative length-scales. In our case we chose $r_0=0.2d_p$, which falls in 
the secondary (dissipative) range, as it can be observed 
from Figs.~\ref{fig:ebsb}-\ref{fig:diff}. 
Note that results do not depend 
on this choice for $0.1<r_0 d_p^{-1}<0.5$ (not shown here). 
In analogy with the single diffusion analysis, we computed the mean squared 
perpendicular particle pair 
separation $\langle r^2\rangle(\tau)$, reported in Fig.~\ref{fig:r2}-(a). 
After an initial transient, the mean square separation manifests a self-similar law, 
with $\langle r^2\rangle\sim \tau^\mu$. The index slightly depends on the plasma beta, 
and is between 1.8 and 2 (see Table I).

Fig.~\ref{fig:r2}-(a) also indicates that after the 
typical separation exceeds the correlation scale $\lambda_c$, normal diffusive behavior is established.
Analogously, the lower boundary is given by the dispersive-dissipative length, here on the order of 
the proton skin depth $d_p$. The vertical arrows represents the characteristic Lagrangian times 
$\tau_{\ell g}$, indicating that the diffusive scaling law for plasmas appears on timescales on the order of this decorrelation mechanism.
Diffusive asymptotic behavior is observed at very large times. 
Lower $\beta$'s show a more clear super-diffusive dispersion, 
while at higher $\beta$ particles are less sensitive to the ${\bf E}\times{\bf B}$ inertial range, which narrows the range of superdiffusion. 
It is evident that the temporal behavior is ``slower'' than the hydrodynamic law in 
Eq.~(\ref{eq:r2}), and this apparent difference will be explained as follows.

In analogy with the Richardson work \cite{Richardson26}, and 
since an exact scaling law for 
compressible anisotropic Vlasov plasmas has not yet been formulated, 
we will study $P(r, \tau)$, namely the probability that particles
are separated by a distance $r$, at a time $\tau$. 
Richardson indeed hypothesized that the probability satisfies \cite{Richardson26,BalkovskyLebedev98}
\beq
\frac{\de P(r,\tau)}{\de \tau}\!=\!\fr{1}{r}\fr{\de }{\de r}\!\left[r\chi(r) \fr{\de P(r, \tau)}{\de r}   \right], ~~\chi(r) = \chi_0 r^{2-\gamma}.
\lab{eq:rich}
\eeq
Here $\chi(r)$ is a scale-dependent eddy-diffusivity due to turbulence, 
and in regular fluids, if the Kolmogorov law is observed, $\chi(r)\sim r^{4/3}$. 
In analogy with his intuition, 
we infer $\chi(r)$ in Eq.~(\ref{eq:rich}) using the exponent in Eq.~(\ref{eq:s2gamma}), 
as suggested by \citet{BalkovskyLebedev98}. 
Given an initial condition $P(r,\tau=0)=\delta(r-r_0)$, and 
$\int\!P(r, \tau)rdr\!=\!1$, Eq.~(\ref{eq:rich}) admits a general solution \cite{BalkovskyLebedev98}:
\beq
P(r, \tau)=\fr{A}{\left( \chi_0 \gamma^2 \tau \right)^{\frac{2}{\gamma}}}e^{-\fr{r^\gamma}{\chi_0 \gamma^2 \tau}} \equiv P_0(\tau)e^{-\fr{r^\gamma}{\chi_0 \gamma^2 \tau}}.
\lab{eq:Pgen}
\eeq
The above is a solution for $r$ sufficiently larger than $r_0$, for separation-times which 
correspond to inertial range length-scales, and where $\gamma$ 
is again the exponent of the second order structure function. 
In Fig.  \ref{fig:r2}-(b), $P(r, \tau)$ is shown, for Run II (all runs have similar results), 
together with Eq.~(\ref{eq:Pgen}).
 The latter have been fitted varying $A$, and keeping the same $\chi_0$ over 
for all the inertial range times. As it can be seen, the distribution describes very well the 
pair dispersion mechanism. In the inset of Fig.~\ref{fig:r2}-(b), the normalized $P(r, \tau)$ 
are reported, rescaling the distribution in time according to Eq.~(\ref{eq:Pgen}). 
The generalized law is clearly observed for intermediate 
times, while is less robust for $\tau\sim26\Omega_{cp}^{-1}$, where $\langle r^2\rangle$ 
approaches $\lambda_C^2$ [compare panel (a) and (b)]. 
Finally, computing moments of Eq.~(\ref{eq:Pgen}),
\beq
\langle r^{\mu \gamma} \rangle \sim \tau^\mu, 
\lab{eq:r2gen}
\eeq
which gives $\mu=2/\gamma$. The latter expectation 
is $\mu\sim1.87$ for Run I and II, and $\sim1.65$ for Run III. These values are in agreement
with the fits of Fig.~\ref{fig:r2}-(a) (see Table I).

Complex diffusive processes have been investigated in 2D plasma turbulence.
In particular, using self-consistent simulations of a hybrid-Vlasov plasma, 
particle diffusion problems have been investigated.
Moderately high resolution simulations have been driven for very long times, in order to 
resolve both short and very long asymptotic behaviors.
The plasma $\beta$ has been varied in order to identify the role of the thermal 
disturbances to the diffusive processes. Particle trajectory show a very interesting and 
complex behavior, being similar to both random walk of magnetic field lines, 
and to test-particles in non-self consistent models of magnetic fields in plasmas \cite{Jokipii66}.
In agreement with fluids, 
the Lagrangian integral time scale  $\tau_\ell$ plays an important role: 
for times much longer than $\tau_\ell$, a classical diffusive 
behavior is observed, with diffusion quantitatively proportional to the plasma beta and 
$\tau_\ell$ inversely 
proportional to $\beta$. For $\tau\ll\tau_\ell$, the particle free-streaming behavior is observed.

For intermediate timescales ($\tau\sim\tau_\ell$), and for inertial range separations,  
particles (and their gyrocenters) undergo superdiffusion, separating very quickly in time
according to Eq.s~(\ref{eq:rich})-(\ref{eq:r2gen}).
The analysis of the probability $P(r,t)$ reveals that dispersion is in agreement with
a generalized Richardson law, depending on the exponent of the spectral index (or the 
exponent of the second-order structure function).  The mean square displacement shows super-diffusive behavior, defined 
by Eq.~(\ref{eq:r2gen}), where $\mu$ is related to the fluctuations scaling.
Results are less pronounced for higher $\beta$ where 
evidently the thermal motion dominates the dispersion and the 
properties of the inertial range are less influential.

  Space plasmas observations and theories suggest than many effects influence the turbulent fluctuations \cite{TesseinEA09,ChandranEA10}, going from strong to weak turbulent regimes. The solutions described by the present numerical experiments, although have been verified here only in few regimes, indicate for the first time that  plasma particles may exhibit a generalized Richardson diffusion. 
The detailed results vary with parameters, e.g., for Kolmogorov scaling 
Eq. (8) would predict $\mu\sim3$, 
while for Iroshnikov-Kraichnan spectra 
it would predict $\mu\sim 4$.
  When this effect is present, bunches of particles undergo a very fast and effective mixing, with the duration of this extraordinary separation being related to the properties of turbulence. The present results must be viewed as a demonstration rather than a universal result, given that, despite covering a wide range of plasma $\beta$, the simulations are restricted to a particular driver, turbulence level, and to 2D. Future work  will extend the above parameters, and explore the role of dimensionality. In 3D, for example, the eddy diffusivity in Eq. (6) may display an anisotropic character, leading to further variations in the Richardson solutions.

This qualitative picture suggests that on the solar corona, for example, 
where more than 4 decades of turbulence are expected, two particles starting at about a proton skin depth 
will depart very quickly, reaching coronal arch sized, very quickly. A similar behavior can be observed 
in general in any space and laboratory plasma, where turbulence can be therefore crucial for heating 
and acceleration processes.


\begin{acknowledgments}
This research is supported in part by NSF Grant No. AGS-1063439 and
AGS-1156094 (SHINE), and by NASA grant NNX14AI63G.
\end{acknowledgments} 


\begin{thebibliography}{33}
\expandafter\ifx\csname natexlab\endcsname\relax\def\natexlab#1{#1}\fi
\expandafter\ifx\csname bibnamefont\endcsname\relax
  \def\bibnamefont#1{#1}\fi
\expandafter\ifx\csname bibfnamefont\endcsname\relax
  \def\bibfnamefont#1{#1}\fi
\expandafter\ifx\csname citenamefont\endcsname\relax
  \def\citenamefont#1{#1}\fi
\expandafter\ifx\csname url\endcsname\relax
  \def\url#1{\texttt{#1}}\fi
\expandafter\ifx\csname urlprefix\endcsname\relax\def\urlprefix{URL }\fi
\providecommand{\bibinfo}[2]{#2}
\providecommand{\eprint}[2][]{\url{#2}}

\bibitem[{\citenamefont{{Kolmogorov}}(1941)}]{Kolmogorov41a}
\bibinfo{author}{\bibfnamefont{A.}~\bibnamefont{{Kolmogorov}}},
  \bibinfo{journal}{Dokl. Akad. Nauk SSSR} \textbf{\bibinfo{volume}{30}},
  \bibinfo{pages}{301} (\bibinfo{year}{1941}).



\bibitem[{\citenamefont{{Langevin}}(1908)}]{Langevin08}
P. Langevin, C. R. Acad. Sci. (Paris) {\bf 146}, 530 (1908); G.~I. Taylor, Proc. London Math. Soc. {\bf 20}, 196 (1921).



\bibitem[{\citenamefont{{Richardson}}(1926)}]{Richardson26}
\bibinfo{author}{\bibfnamefont{L.~F.} \bibnamefont{{Richardson}}},
  \bibinfo{journal}{Proc. R. Soc. A}
  \textbf{\bibinfo{volume}{110}}, \bibinfo{pages}{709} (\bibinfo{year}{1926}). 

\bibitem[{\citenamefont{{Taylor} and {McNamara}}(1971)}]{TaylorMcNamara71}
\bibinfo{author}{\bibfnamefont{J.~B.} \bibnamefont{{Taylor}}} \bibnamefont{and}
  \bibinfo{author}{\bibfnamefont{B.}~\bibnamefont{{McNamara}}},
  \bibinfo{journal}{Phys. Fluids} \textbf{\bibinfo{volume}{14}},
  \bibinfo{pages}{1492} (\bibinfo{year}{1971}). 


\bibitem[Hauff et al.(2009)]{HauffEA09}
T. Hauff {\it et al.}, Phys. Rev. Lett. {\bf 102}, 075004 (2009).


\bibitem[{\citenamefont{{Jokipii} and {Parker}}(1969)}]{JokipiiParker69}
\bibinfo{author}{\bibfnamefont{J.~R.} \bibnamefont{{Jokipii}}}
  \bibnamefont{and} \bibinfo{author}{\bibfnamefont{E.~N.}
  \bibnamefont{{Parker}}}, \bibinfo{journal}{\apj}
  \textbf{\bibinfo{volume}{155}}, \bibinfo{pages}{777} (\bibinfo{year}{1969}).






\bibitem[{\citenamefont{{Eyink} et~al.}(2013)\citenamefont{{Eyink}, {Vishniac},
  {Lalescu}, {Aluie}, {Kanov}, {B{\"u}rger}, {Burns}, {Meneveau}, and
  {Szalay}}}]{EyinkEA13}
\bibinfo{author}{\bibfnamefont{G.}~\bibnamefont{{Eyink} {\it et al.}}},
  \bibinfo{journal}{Nature} \textbf{\bibinfo{volume}{497}}, \bibinfo{pages}{466}
  (\bibinfo{year}{2013}). 

\bibitem[{\citenamefont{{Lazarian} et~al.}(2015)\citenamefont{{Lazarian},
  {Eyink}, {Vishniac}, and {Kowal}}}]{LazarianEA15}
\bibinfo{author}{\bibfnamefont{A.}~\bibnamefont{{Lazarian} {\it et al.}}},
  \bibinfo{journal}{Phil. Trans. R. Soc. A} \textbf{\bibinfo{volume}{373}}, \bibinfo{pages}{20140144}
  (\bibinfo{year}{2015}). 



\bibitem[{\citenamefont{{Lepreti} et~al.}(2012)\citenamefont{{Lepreti},
  {Carbone}, {Abramenko}, {Yurchyshyn}, {Goode}, {Capparelli}, and
  {Vecchio}}}]{LepretiEA12}
\bibinfo{author}{\bibfnamefont{F.}~\bibnamefont{{Lepreti} {\it et al.}}},
  \bibinfo{journal}{Astrophys. J. Lett.} \textbf{\bibinfo{volume}{759}},
  \bibinfo{eid}{L17} (\bibinfo{year}{2012}). 

\bibitem[{\citenamefont{{Ruffolo} et~al.}(2004)\citenamefont{{Ruffolo},
  {Matthaeus}, and {Chuychai}}}]{RuffoloEA04}
\bibinfo{author}{\bibfnamefont{D.}~\bibnamefont{{Ruffolo}}},
  \bibinfo{author}{\bibfnamefont{W.~H.} \bibnamefont{{Matthaeus}}},
  \bibnamefont{and}
  \bibinfo{author}{\bibfnamefont{P.}~\bibnamefont{{Chuychai}}},
  \bibinfo{journal}{\apj} \textbf{\bibinfo{volume}{614}}
  (\bibinfo{year}{2004}).

\bibitem[{\citenamefont{{Ruffolo} et~al.}(2003)\citenamefont{{Ruffolo},
  {Matthaeus}, and {Chuychai}}}]{RuffoloEA03}
\bibinfo{author}{\bibfnamefont{D.}~\bibnamefont{{Ruffolo}}},
  \bibinfo{author}{\bibfnamefont{W.~H.} \bibnamefont{{Matthaeus}}},
  \bibnamefont{and}
  \bibinfo{author}{\bibfnamefont{P.}~\bibnamefont{{Chuychai}}},
  \bibinfo{journal}{Astrophys. J. Lett.} \textbf{\bibinfo{volume}{597}},
  \bibinfo{pages}{L169} (\bibinfo{year}{2003}).
 

\bibitem[Perrone et al.(2013)]{PerroneEA13}
D. Perrone {\it et al.}, Space Sci. Rev. {\bf 178}, 233 (2013). 

\bibitem[{\citenamefont{{Jokipii}}(1966)}]{Jokipii66}
\bibinfo{author}{\bibfnamefont{J.~R.} \bibnamefont{{Jokipii}}},
  \bibinfo{journal}{Astrophys. J.} \textbf{\bibinfo{volume}{146}},
  \bibinfo{pages}{480} (\bibinfo{year}{1966}).



\bibitem[Balescu et al.(1994)]{Balescu94}
R. Balescu, H.-D. Wang, and J. H. Misguich, Phys. Plasmas {\bf 1}, 3826 (1994).

\bibitem[Busse et al.(2007)]{BusseEA07}
A. Busse, W.-C. M{\"u}ller, H. Homann, and R. Grauer, Phys. Plasmas {\bf 14}, 122303 (2007).

\bibitem[{\citenamefont{{Dmitruk} et~al.}(2004)\citenamefont{{Dmitruk},
  {Matthaeus}, and {Seenu}}}]{DmitrukEA04}
\bibinfo{author}{\bibfnamefont{P.}~\bibnamefont{{Dmitruk}}},
  \bibinfo{author}{\bibfnamefont{W.~H.} \bibnamefont{{Matthaeus}}},
  \bibnamefont{and} \bibinfo{author}{\bibfnamefont{N.}~\bibnamefont{{Seenu}}},
  \bibinfo{journal}{Astrophys. J.} \textbf{\bibinfo{volume}{617}},
  \bibinfo{pages}{667} (\bibinfo{year}{2004}). 



\bibitem[{\citenamefont{{Zimbardo} et~al.}(2006)\citenamefont{{Zimbardo},
  {Pommois}, and {Veltri}}}]{ZimbardoEA06}
\bibinfo{author}{\bibfnamefont{G.}~\bibnamefont{{Zimbardo}}},
  \bibinfo{author}{\bibfnamefont{P.}~\bibnamefont{{Pommois}}},
  \bibnamefont{and} \bibinfo{author}{\bibfnamefont{P.}~\bibnamefont{{Veltri}}},
  \bibinfo{journal}{Astrophys. J. Lett.} \textbf{\bibinfo{volume}{639}},
  \bibinfo{pages}{L91} (\bibinfo{year}{2006}).



\bibitem[Marsch(1990)]{Marsch06}
E. Marsch, Liv. Rev. Solar Phys. {\bf 3}, 1 (2006).



\bibitem[{\citenamefont{{Batchelor}}(1950)}]{Batchelor50}
\bibinfo{author}{\bibfnamefont{G.~K.} \bibnamefont{{Batchelor}}},
  \bibinfo{journal}{Q. J. R. Meteorol. Soc.}
  \textbf{\bibinfo{volume}{76}}, \bibinfo{pages}{133} (\bibinfo{year}{1950}).

\bibitem[{\citenamefont{{Falkovich} et~al.}(2001)\citenamefont{{Falkovich},
  {Gaw{\c e}dzki}, and {Vergassola}}}]{FalkovichEA01}
\bibinfo{author}{\bibfnamefont{G.}~\bibnamefont{{Falkovich}}},
  \bibinfo{author}{\bibfnamefont{K.}~\bibnamefont{{Gaw{\c e}dzki}}},
  \bibnamefont{and}
  \bibinfo{author}{\bibfnamefont{M.}~\bibnamefont{{Vergassola}}},
  \bibinfo{journal}{Rev. Mod. Phys.} \textbf{\bibinfo{volume}{73}},
  \bibinfo{pages}{913} (\bibinfo{year}{2001}). 




\bibitem[{\citenamefont{{Servidio} et~al.}(2012)
\citenamefont{{Servidio}, {Valentini}, {Califano}, and {Veltri}}}]{ServidioEA12}
\bibinfo{author}{\bibfnamefont{S.}~\bibnamefont{{Servidio} {\it et al.}}},
\bibinfo{journal}{Phys. Rev. Lett.} \textbf{\bibinfo{volume}{108}},
\bibinfo{eid}{045001} (\bibinfo{year}{2012}). 

\bibitem[{\citenamefont{{Haynes} et~al.}(2014)\citenamefont{{Haynes},
  {Burgess}, and {Camporeale}}}]{HaynesEA14}
\bibinfo{author}{\bibfnamefont{C.~T.} \bibnamefont{{Haynes}}},
  \bibinfo{author}{\bibfnamefont{D.}~\bibnamefont{{Burgess}}},
  \bibnamefont{and}
  \bibinfo{author}{\bibfnamefont{E.}~\bibnamefont{{Camporeale}}},
  \bibinfo{journal}{\apj} \textbf{\bibinfo{volume}{783}}, \bibinfo{eid}{38}
  (\bibinfo{year}{2014}). 

\bibitem[{\citenamefont{{Franci} et~al.}(2015)\citenamefont{{Franci}, {Landi},
  {Matteini}, {Verdini}, and {Hellinger}}}]{FranciEA15}
\bibinfo{author}{\bibfnamefont{L.}~\bibnamefont{{Franci} {\it et al.}}},
  \bibinfo{journal}{\apj} \textbf{\bibinfo{volume}{812}}, \bibinfo{eid}{21}
  (\bibinfo{year}{2015}). 















\bibitem[{\citenamefont{{Winske}}(1985)}]{Winske85}
\bibinfo{author}{\bibfnamefont{D.}~\bibnamefont{{Winske}}},
  \bibinfo{journal}{Space Sci. Rev.} \textbf{\bibinfo{volume}{42}},
  \bibinfo{pages}{53} (\bibinfo{year}{1985}). 

\bibitem[{\citenamefont{{Matthews}}(1994)}]{Matthews94}
\bibinfo{author}{\bibfnamefont{A.~P.} \bibnamefont{{Matthews}}},
  \bibinfo{journal}{J. Comp. Phys.}
  \textbf{\bibinfo{volume}{112}}, \bibinfo{pages}{102} (\bibinfo{year}{1994}).

\bibitem[{\citenamefont{{Chen} et~al.}(1993)\citenamefont{{Chen}, {Doolen},
  {Kraichnan}, and {She}}}]{ChenEA93}
\bibinfo{author}{\bibfnamefont{S.}~\bibnamefont{{Chen} {\it et al.}}},
  \bibinfo{journal}{Phys. Fluids} \textbf{\bibinfo{volume}{5}},
  \bibinfo{pages}{458} (\bibinfo{year}{1993}).

\bibitem[{\citenamefont{{Bec} et~al.}(2010)\citenamefont{{Bec}, {Biferale},
  {Lanotte}, {Scagliarini}, and {Toschi}}}]{BecEA10}
\bibinfo{author}{\bibfnamefont{J.}~\bibnamefont{{Bec} {\it et al.}}},
  \bibinfo{journal}{J. Fluid Mech.} \textbf{\bibinfo{volume}{645}},
  \bibinfo{pages}{497} (\bibinfo{year}{2010}), \eprint{0904.2314}.

\bibitem[{\citenamefont{{Thalabard} et~al.}(2014)\citenamefont{{Thalabard},
  {Krstulovic}, and {Bec}}}]{ThalabardEA14}
\bibinfo{author}{\bibfnamefont{S.}~\bibnamefont{{Thalabard}}},
  \bibinfo{author}{\bibfnamefont{G.}~\bibnamefont{{Krstulovic}}},
  \bibnamefont{and} \bibinfo{author}{\bibfnamefont{J.}~\bibnamefont{{Bec}}},
  \bibinfo{journal}{J. Fluid Mech.} \textbf{\bibinfo{volume}{755}},
  \bibinfo{eid}{R4} (\bibinfo{year}{2014}).

\bibitem[{\citenamefont{{Mininni} and {Pouquet}}(2009)}]{Mininni09}
\bibinfo{author}{\bibfnamefont{P.~D.} \bibnamefont{{Mininni}}}
  \bibnamefont{and}
  \bibinfo{author}{\bibfnamefont{A.}~\bibnamefont{{Pouquet}}},
  \bibinfo{journal}{\pre} \textbf{\bibinfo{volume}{80}}, \bibinfo{eid}{025401}
  (\bibinfo{year}{2009}).

\bibitem[{\citenamefont{{Lee} et~al.}(2010)\citenamefont{{Lee}, {Brachet},
  {Pouquet}, {Mininni}, and {Rosenberg}}}]{LeeEA10}
\bibinfo{author}{\bibfnamefont{E.}~\bibnamefont{{Lee} {\it et al.}}},
  \bibinfo{journal}{\pre} \textbf{\bibinfo{volume}{81}}, \bibinfo{eid}{016318}
  (\bibinfo{year}{2010}).

\bibitem[{\citenamefont{{Tessein} et~al.}(2009)\citenamefont{{Tessein},
  {Smith}, {MacBride}, {Matthaeus}, {Forman}, and {Borovsky}}}]{TesseinEA09}
\bibinfo{author}{\bibfnamefont{J.~A.} \bibnamefont{{Tessein} {\it et al.}}},
\bibinfo{journal}{\apj}
  \textbf{\bibinfo{volume}{692}}, \bibinfo{pages}{684} (\bibinfo{year}{2009}).


\bibitem[{\citenamefont{{Tooprakai} et~al.}(2007)\citenamefont{{Tooprakai},
  {Chuychai}, {Minnie}, {Ruffolo}, {Bieber}, and {Matthaeus}}}]{TooprakaiEA07}
\bibinfo{author}{\bibfnamefont{P.}~\bibnamefont{{Tooprakai} {\it et al.}}},
\bibinfo{journal}{Geophys. Res. Lett.}
  \textbf{\bibinfo{volume}{34}}, \bibinfo{pages}{17105} (\bibinfo{year}{2007}).

\bibitem[{\citenamefont{{Drake} et~al.}(2010)\citenamefont{{Drake}, {Opher},
  {Swisdak}, and {Chamoun}}}]{DrakeEA10}
\bibinfo{author}{\bibfnamefont{J.~F.} \bibnamefont{{Drake} {\it et al.}}}, 
\bibinfo{journal}{\apj}
  \textbf{\bibinfo{volume}{709}}, \bibinfo{pages}{963} (\bibinfo{year}{2010}).

\bibitem[{\citenamefont{{Zank} et~al.}(2015)\citenamefont{{Zank}, {Hunana},
  {Mostafavi}, {Le Roux}, {Li}, {Webb}, {Khabarova}, {Cummings}, {Stone}, and
  {Decker}}}]{ZankEA15}
\bibinfo{author}{\bibfnamefont{G.~P.} \bibnamefont{{Zank} {\it et al.}}},
  \bibinfo{journal}{\apj} \textbf{\bibinfo{volume}{814}}, \bibinfo{eid}{137}
  (\bibinfo{year}{2015}).

\bibitem[{\citenamefont{{Seripienlert}
  et~al.}(2010)\citenamefont{{Seripienlert}, {Ruffolo}, {Matthaeus}, and
  {Chuychai}}}]{SeripienlertEA10}
\bibinfo{author}{\bibfnamefont{A.}~\bibnamefont{{Seripienlert} {\it et al.}}},
  \bibinfo{journal}{\apj} \textbf{\bibinfo{volume}{711}}, \bibinfo{pages}{980}
  (\bibinfo{year}{2010}).

\bibitem[{\citenamefont{{Okuda} and {Dawson}}(1973)}]{OkudaDawson73}
\bibinfo{author}{\bibfnamefont{H.}~\bibnamefont{{Okuda}}} \bibnamefont{and}
  \bibinfo{author}{\bibfnamefont{J.~M.} \bibnamefont{{Dawson}}},
  \bibinfo{journal}{Phys. Fluids} \textbf{\bibinfo{volume}{16}},
  \bibinfo{pages}{408} (\bibinfo{year}{1973}).

\bibitem[{\citenamefont{{Balkovsky} and {Lebedev}}(1998)}]{BalkovskyLebedev98}
\bibinfo{author}{\bibfnamefont{E.}~\bibnamefont{{Balkovsky}}} \bibnamefont{and}
  \bibinfo{author}{\bibfnamefont{V.}~\bibnamefont{{Lebedev}}},
  \bibinfo{journal}{\pre} \textbf{\bibinfo{volume}{58}}, \bibinfo{pages}{5776}
  (\bibinfo{year}{1998}).



\bibitem[{\citenamefont{{Chandran} et~al.}(2010)\citenamefont{{Chandran}, {Li},
  {Rogers}, {Quataert}, and {Germaschewski}}}]{ChandranEA10}
\bibinfo{author}{\bibfnamefont{B.~D.~G.} \bibnamefont{{Chandran} {\it et al.}}},
  \bibinfo{journal}{\apj} \textbf{\bibinfo{volume}{720}}, \bibinfo{pages}{503}
  (\bibinfo{year}{2010}).

\end{thebibliography}

\end{document}